\documentstyle[12pt]{article}
\newcommand{\slp}{\raise.15ex\hbox{$/$}\kern-.57em\hbox{$\partial$}}
\newcommand{\sla}{\raise.15ex\hbox{$/$}\kern-.57em\hbox{$a$}}
\newcommand{\slA}{\raise.15ex\hbox{$/$}\kern-.57em\hbox{$A$}}
\newcommand{\slb}{\raise.15ex\hbox{$/$}\kern-.57em\hbox{$b$}}
\newcommand{\be}{\begin{equation}}

\newcommand{\ee}{\end{equation}}
\newcommand{\bear}{\begin{eqnarray}}
\def\eear{\end{eqnarray}}
\newcommand{\ear}{\end{eqnarray}}
\newcommand{\ba}{\begin{eqnarray*}}
\newcommand{\ea}{\end{eqnarray*}}
\begin{document}
\begin{titlepage}
\setcounter{page}{1}
\begin{flushright}
HD--THEP--99--24\\
\end{flushright}
\vskip1.5cm
\begin{center}
{\large{\bf Master Equation for Lagrangian Gauge Symmetries}}\\
\vspace{1cm}
R. Banerjee, 
%%%%%
\footnote{email: r.banerjee@thphys.uni-heidelberg.de\\
On leave of absence from S.N. Bose Natl. Ctr. for Basic Sc., Salt Lake, 
Calcutta 700091, India }
%%%%%
H.J. Rothe 
%%%%%%
\footnote{email: h.rothe@thphys.uni-heidelberg.de}
%%%%%%
and 
K. D. Rothe 
%%%%%%%
\footnote{email: k.rothe@thphys.uni-heidelberg.de}
%%%%%%
\\
{\it Institut  f\"ur Theoretische Physik - Universit\"at Heidelberg}
\\
{\it Philosophenweg 16, D-69120 Heidelberg, Germany}

{(July 27, 1999)}
\end{center}

\begin{abstract}
\noindent
Using purely Hamiltonian methods we derive a simple differential equation 
for the generator of the most general local symmetry transformation of 
a Lagrangian. The restrictions on the gauge parameters found by earlier 
approaches are easily reproduced from this equation. We also discuss the connection with the purely
Lagrangian approach. The general considerations are applied to the Yang-Mills 
theory. 
\end{abstract}
\end{titlepage}
%%%%%%%%%%%%%%%%%%%%%%%%%%%%%%%%%%%%%%%%%%%%%%%%%%%%%%%%%%%%%%%%%%%%%%%%%%%
%%%%%%%%%%%%%%%%%%%%%%%%%%%%%%%%%%%%%%%%%%%%%%%%%%%%%%%%%%%%%%%%%%%%%%%%%%%%

\newpage

The problem of finding the most general local symmetries of a Lagrangian
has been pursued by various authors, using either Lagrangian 
\cite{Mukunda,Gitman,Chaichian1,Shirzad1} or Hamiltonian techniques \cite{Henneaux,Chaichian,Wipf}. Nevertheless a 
compact equation which determines the precise structure of the generator of gauge transformations, which are the symmetries of a Lagrangian, is still lacking. 

In a recent paper \cite{Banerjee} we had shown that the requirement of commutativity
of the time derivative operation with an arbitrary infinitesimal gauge transformation generated by the first class constraints was the only input needed
for obtaining the restrictions on the gauge parameters 
entering the most general form of the generator $G$. 
The analysis was performed entirely in the
Hamiltonian framework.

Following essentially the same commutativity requirement, 
we derive here, in a Hamiltonian framework, a simple differential equation for the generator. This differential equation encodes, in particular, the restrictions on the gauge parameters. We explicitly demonstrate that
this equation implies the (off shell) invariance of the action under
the transformation generated by $G$, and ensures the covariance of the 
Hamilton equations of motion. 

In this paper we shall consider purely first class systems. The extension to mixed first and second class systems is 
straightforward. To keep the algebra simple, and also for reasons of comparison, we assume all constraints to be irreducible.

We consider a Hamiltonian system whose dynamics is described by the
total Hamiltonian
\be\label{totalhamiltonian}
H_T = H_c + \sum_{a_1} v^{a_1}\Phi_{a_1}\;.
\ee
where $H_c$ is the canonical Hamiltonian, $\{\Phi_{a_1}\approx 0\}$ are the 
(first class) primary constraints, and $v^{a_1}$ are the associated
Lagrange multipliers. We denote the complete set of 
(primary and secondary) constraints
%%%
\footnote{``Secondary" refers to all generations of constraints beyond the primary one.}
%%%
by
$\{\Phi_a\} = \{\Phi_{a_1},\Phi_{a_2}\}$.
The Poisson algebra of the constraints with themselves
and with the canonical Hamiltonian, is of the form
\be\label{algebra}
[H_c,\Phi_a] = V_a^b \Phi_b\label{algebra1}
\ee
\be
[\Phi_a,\Phi_b] = C_{ab}^c\Phi_c\label{algebra}
\ee
where $V_a^b$ and $C_{ab}^c$ may be functions of the 
phase-space variables.
Consider an infinitesimal transformation on the coordinates generated by $G$,
\be\label{delta}
\delta q^i = [q^i,G]\;,
\ee
with 
\be\label{generator}
G = \sum_a \epsilon^a \Phi_a
\ee
where, following Dirac's conjecture \cite{Dirac}, the sum includes 
all of the first class constraints. The gauge parameters 
are allowed to depend in general on time, as well as on the 
phase space variables and Lagrange multipliers.

In (\ref{delta}) we have chosen to include the gauge parameters $\epsilon^a$ inside the bracket. The difference between (\ref{delta}) and the 
variation $\delta q^i$ computed with the parameters outside the 
bracket is proportional to the constraints. Such terms can always be 
be written in the form $\Lambda_{ij}\frac{\delta S}{\delta q_j}$, with 
$\Lambda_{ij} = -\Lambda_{ji}$, which correspond to trivial gauge transformations \cite{Henneaux}. In this paper we are only considering 
gauge transformations, modulo these trivial gauge transformations.  

Consider the first of Hamilton's equations, giving the connection between the
velocities and the momenta 
\be\label{total}
\frac{dq^i}{dt} \approx [q^i,H_T]\;
\ee
where $\approx$ stands for ``weak equality" in the sense of Dirac \cite{Dirac}.
>From above we obtain
\be\label{dotdeltaq}
\frac{d}{dt}\delta q^i \approx [[q^i,G],H_T] + 
[q^i,\frac{\partial}{\partial t}G]\;,
\ee
and 
\be\label{deltaqdot}
\delta \frac{dq^i}{dt}\approx [[q^i,H_T],G]\;.
\ee
where we have taken account of the fact that $G$ will in general depend 
exlicitly  on the time.
Implementing the commutativity requirement \cite{Banerjee}
\be\label{commutativity}
\frac{d}{dt}\delta q^i =  \delta\frac{d}{dt}q^i
\ee
by equating the last two expressions, and using the Jacobi identity, we arrive at the condition
\be\label{condition1}
[q^i,\frac{\partial}{\partial t}G + [G,H_T]] \approx 0\;
\ee
Using the ansatz (\ref{generator}) as well as the algebra given in eqs. (\ref{algebra1}) and (\ref{algebra}), it follows that
$\frac{\partial}{\partial t}G + [G,H_T]$ is given by a linear combination
of the first class constraints:
\be\label{sum}
\frac{\partial}{\partial t}G + [G,H_T] = \sum_a \xi^a \Phi_a\;.
\ee
Substituting this expression into (\ref{condition1}), we arrive at
the condition 
\be\label{conditiononxi}
\xi^a \frac{\partial}{\partial p_i}\Phi_a \approx 0\;.
\ee
Now, the first class nature and linear independence (irreducibility) of
the constraints guarantees that each of these can be identified as a momentum conjugate to some coordinate, the precise mapping being effected by a canonical transformation \cite{Gitman}. Since (\ref{conditiononxi}) holds for all $i$ one is led to the condition
$\xi^a \approx 0$. Therefore the r.h.s of (\ref{sum}) will be proportional 
to the square of the constraints, so that, within the
Hamiltonian formalism, we are allowed to set the l.h.s of 
(\ref{sum}) strongly equal to zero:
\be\label{master}
\frac{\partial}{\partial t}G + [G,H_T] = 0\;.
\ee
This is the fundamental equation determining $G$, which we henceforth refer to as the ``master equation". As we shall see, it
will guarantee the (off-shell) invariance of the total action.

The condition (\ref{master}) also ensures the covariance of the Hamilton
equations of motion under a transformation generated by $G$. Thus consider
the equation of motion for $q^i$:
\be\label{eqmotion}
\frac{dq^i}{dt} \approx [q^i,H_T(q,p,v)]\;.
\ee
Consider further the gauge-transfomed phase space variables
and Lagrange multipliers
\[
\bar q^i = q^i + \delta q^i\;,\quad \bar p_i = p_i + \delta p_i\;, \quad
\bar v^{a_1} = v^{a_1} + \delta v^{a_1}
\]
with
\[
\delta q^i = [q^i,G] \;,\quad \delta p_i = [p_i,G]\;,\quad \delta v^{a_1} = 
[v^{a_1},G]\ .
\]
Using the equations of motion (\ref{eqmotion}), the master equation 
(\ref{master}) as well as (\ref{dotdeltaq})
one readily verifies that 
\[ 
\frac{d{\bar q}_i}{dt} \approx [\bar q^i,H_T(\bar q, \bar p, \bar v)]\;,
\]
which demonstrates the covariance. A similar statement holds for 
$\bar p_i$.

We now examine the implications of our condition  (\ref{master})
for the gauge parameters in (\ref{generator}). Making use of the algebra (\ref{algebra1}) and (\ref{algebra}), the master equation
(\ref{master}) is easily seen to lead to
\be
\frac{\partial}{\partial t}G + [G,H_T]=
\Big(\frac{d\epsilon^b}{dt}   - \epsilon^a 
[V_a^b  + v^{a_1} C_{a_1a}^b]\Big)\Phi_b - \delta v^{a_1} \Phi_{a_1}
 = 0\;.\nonumber
\ee
>From here we obtain the following conditions
\bear
\delta v^{b_1}&=& \frac{d\epsilon^{b_1}}{dt}   - \epsilon^a 
[V_a^{b_1}  +  v^{a_1} C_{a_1a}^{b_1}]\;,\label{multiplier}\\
 0&=&\frac{d\epsilon^{b_2}}{dt}  - \epsilon^a 
[V_a^{b_2}  +  v^{a_1} C_{a_1a}^{b_2}]\;. \label{parameter}
\eear
Note that in the above equations, $\frac{d\epsilon^a}{dt}$ denotes the total time
derivative as given by
\be
\frac{d\epsilon^a}{dt} = \frac{D\epsilon^a}{Dt} + [\epsilon^a,H_c]+
v^{a_1}[\epsilon^a,\Phi_{a_1}]\;
\ee
where, following the notation of Ref. \cite{Henneaux}, 
\be\label{Dderivative}
\frac{D}{Dt} = \frac{\partial}{\partial t}
+ \dot v^{a_1}\frac{\partial}{\partial v^{a_1}} + 
{\ddot v}^{a_1}\frac{\partial}{\partial {\dot v}^{a_1}} + \cdot\cdot\cdot.
\ee
with an overdot denoting the derivative with respect 
to the explicit dependence
in time.
The same conditions have been obtained in Ref. \cite{Henneaux}  by looking at the invariance of the gauge-fixed {\it extended} action,
 and directly from the commutativity requirement (\ref{commutativity})
 in Ref. \cite{Banerjee}. Note that
equation (\ref{multiplier}) only plays a role in the Hamiltonian formulation where 
the equations of motion are obtained from the variation of the total 
action. For the extraction of the symmetries of the original Lagrangian the relevant equation is (\ref{parameter}).

It is clear that the above considerations can be easily extended to the case where the dynamics is described by the {\it extended} Hamiltonian
$H_E = H_T + v^{a_2}\Phi_{a_2}$. The commutativity requirement
will now lead to the extended master equation
\be\label{master2}
\frac{\partial}{\partial t}G + [G,H_E] = 0\;.
\ee
In this case no restrictions on the gauge parameters are implied by
eq. (\ref{master2}), which only determines the transformation law
for the multipliers $v^a$:
\be
\Big(\frac{d\epsilon^b}{dt}   - \epsilon^a 
[V_a^b  + v^{c} C_{ca}^b]\Big) - \delta v^{b} 
 = 0\;.\nonumber
\ee
This equation was obtained in ref. \cite{Henneaux} by directly looking at the
invariance of the extended action. Furthermore, by imposing gauge conditions 
implementing $\{v^{a_2}=0\}$
\cite{Henneaux}, one recovers (\ref{multiplier}) and
(\ref{parameter}), as is evident from (\ref{master2}). 

Returning to our formulation
in terms of the total Hamiltonian,
the first step for obtaining the final form for $G$ consists 
in solving equation  (\ref{parameter}) for the $\epsilon^{a_2}$'s 
in terms of the coordinates, momenta, Lagrange multiplieres (including their 
time derivatives) and a set of independent parameters
whose number equals the number of primary constraints. These parameters can be taken to be a function 
of time only. A method for solving these equations has been given
in \cite{Henneaux}. The final step consists in computing the variations 
(\ref{delta}), and in eliminating the canonical momenta and the 
Lagrange multipliers in terms of the coordinates and velocities using the  
first of the Hamilton equations of motion. In particular the multipliers are eliminated by 
making use of the Hamilton equations for the variables which are conjugate to the primary constraints. In fact equation (\ref{multiplier}) is just a consistency condition of the entire scheme, as we now show.

The primary constraints can always be expressed in the form
\cite{Mukunda}
\be\label{primaryconstraints}
\Phi_{a_1} = p_{a_1} - f_{a_1}(\{q^a\},\{p_{a_2}\})\;,
\ee
where $q^{a_1},p_{a_1}$ are canonically conjugate pairs of variables,
with $\{\dot q ^{a_1}\}$  the (arbitrary) non projectible velocities. Taking 
the variation of 
\be
\frac{dq^{a_1}}{dt} \approx [q^{a_1},H_c] +[q^{a_1},\Phi_{b_1}] v^{b_1}
\ee
we have
\be
\delta\frac{d q^{a_1}}{dt} \approx \delta[q^{a_1},H_c] + [q^{a_1},\Phi_{b_1}]\delta v^{b_1} 
+ \delta[q^{a_1},\Phi_{b_1}] v^{b_1}\;.
\ee
Using (\ref{commutativity}) we obtain for the l.h.s,
\be
\delta\frac{dq^{a_1}}{dt} = \frac{d}{dt}\delta q^{a_1} \approx 
\frac{d\epsilon^{a}}{dt} [q^{a_1},\Phi_a] + 
\epsilon^a\left([[q^{a_1},\Phi_a],H_c] + 
[[q^{a_1},\Phi_a],\Phi_{c_1}]v^{c_1}\right)\;.
\ee
Making use of the Jacobi identity as well as of
(\ref{parameter}) , one readily finds
\be\label{weakrelations}
\left(\delta v^{b_1}-\frac{d\epsilon^{b_1}}{dt}   +\epsilon^a 
[V_a^{b_1}  +  v^{a_1} C_{a_1a}^{b_1}]\right)[q^{a_1},\Phi_{b_1}]\approx 0\;.
\ee
Recalling (\ref{primaryconstraints}), and noting that
the gauge transformations are defined only modulo the trivial ones,
we make use of this freedom in order to obtain from (\ref{weakrelations}) the strong relations (\ref{multiplier}).

To make contact with previous literature, we recall  
the conditions given in ref.\cite{Pons,Gomis},
%%%%%%%
\footnote{In ref. \cite{Pons}, eq.(\ref{Ponsconditions2})
was given in the form  $[G,\Phi_{a_1}] = g_{a_1}^{b_1} \Phi_{b_1}$.}
%%%%%%
\bear
&&[H_c,G] - \frac{\partial G}{\partial t} = h^{a_1} \Phi_{a_1}\label{Ponsconditions1}\\
&&[\Phi_{a},\Phi_{b_1}] = C_{a b_1}^{c_1} \Phi_{c_1}\;.\label{Ponsconditions2}
\eear
where $G$ 
is of course always understood to be first-class. As was however emphasized in 
ref \cite{Henneaux}, the second condition is restrictive.
Indeed, if we take equations
(\ref{Ponsconditions2}) together with  the Ansatz
(\ref{generator}) as our starting point, we are led to
\be\label{Shirzadeqn}
 0= \frac{d\epsilon^{b_2}}{dt}  - \epsilon^a V_a^{b_2}\;.
\ee
as the only condition.
Note that this condition follows from our 
general relations (\ref{multiplier}) and
(\ref{parameter}),
since the structure functions $C_{a_1,a}^{b_2}$ 
are assumed to be zero, as implied by assumption  
(\ref{Ponsconditions2}). Note also that our first
relation is absent since their conditions do not involve any
Lagrange multipliers.

We now want to make contact with the purely Lagrangian methods of obtaining 
the gauge symmetries \cite{Mukunda,Gitman,Chaichian1,Shirzad1}. 
As discussed by Dirac \cite{Dirac}, the classical Euler-Lagrange equations follow 
from the action principle $\delta S_T=0$, where $S_T$ is defined by
\be\label{action}
S_T = \int dt [p_i\dot q^i - H_T]\;.
\ee
We now show that the condition (\ref{master}) does indeed ensure the invariance of the
total action under the transformations generated by $G$.
Consider
$L_T = p_i\dot q^i - H_T$. Assuming the commutativity (\ref{commutativity}),
we find for an infinitesimal transformation generated by $G$
\bear
\delta L_T &=& [p_i,G]\dot q^i - \dot p_i [q^i,G] -[H_T,G] + 
\frac{d}{dt}(p_i\delta q^i)\nonumber\\
&=& \frac{\partial}{\partial t}G + [G,H_T]+\frac{d}{dt}(-G + p_i\delta q^i)\;.
\eear
Since the endpoint configurations in the total action (\ref{action}) are taken to be fixed, we see that the invariance of the total action
under this transformation leads to the off-shell condition 
(\ref{master}). Observe that no use has been made of the equations of motion.

The general 
variation of the Lagrangian $L(q,\dot q)$ 
is given by 
\be
\delta L = -L_i\delta q^i + \frac{d}{dt}\Big(\frac{\partial L}
{\partial{\dot q^i}}\delta q^i\Big)
\ee
where $L_i$ is the Euler derivative, given in terms of the Hessian $W_{ij}$ by
\be
L_i = W_{ij}\ddot q^j + 
\frac{\partial^2L}{\partial q^j\partial\dot q^i}\dot q^j -
\frac{\partial L}{\partial q^i}
\ee
Note that $-L_i\delta q^i$ in the present formulation corresponds to
$\frac{\partial}{\partial t}G + [G,H_T]-\frac{d}{dt}G$ in the formulation
in terms of the total action. Both expressions vanish on shell.
 
It is well known \cite{Mukunda,Gitman,Chaichian1,Shirzad1} that to each 
gauge symmetry of the Lagrangian there  is a corresponding gauge 
identity having the general form
\be\label{gaugeidentities}
\Lambda_{a_1} = \sum_k \frac{d^k}{dt^k}(\rho^{i}_{(k)a_1}(q,\dot q)L_i) 
\equiv 0
\ee
Taking a general gauge transformation of the form \cite{Gitman}
%%%%%%%
\footnote{The form of $\delta q^i$ as given by 
(\ref{deltaq}) is also suggested within the Hamiltonian framework 
by the work of ref. \cite{Henneaux}, with $\eta^{a_1}$ playing the role of the
independent gauge parameters.}
%%%%%%% 
\be\label{deltaq}
\delta q^i = \sum_{k,a_1}(-1)^k\frac{d^k\eta^{a_1}(t)}{dt^k}
\rho^{i}_{(k)a_1}(q,\dot q)
\ee
where $\eta^{a_1}(t)$ are the gauge parameters, one finds that the variation of the Lagrangian is given by 
\be
\delta L = -\Lambda_{a_1}\eta^{a_1} + \cdot\cdot\cdot
\ee
where the ``dots" stand for a contribution given by a total time derivative, which does not contribute to the variation of the action. Because of the gauge identities the 
action is invariant. The corresponding statement in the case of 
the total action (\ref{action}) is that, once the equations (\ref{parameter}) are solved, 
the master equation (\ref{master}) is satisfied identically 
without making use of 
the equations of motion. From the 
above point of view the difficulty in solving the eqs. (\ref{parameter}) 
manifests itself in the problem of finding the zero modes of the 
Hessian, from which the functions $\rho^{i}_{(k)a_1}$ in 
(\ref{gaugeidentities}) are determined.

The number of gauge identities is equal to the number of independent gauge
parameters. This also coincides with the number of first class primary
constraints. Thus, for Hamiltonian systems involving only such
constraints, there are no restrictions on the parameters parametrizing
the gauge generators (\ref{generator}). An important class of
systems having this property are the so called ``zero-Hamiltonian" systems
where $H_c\equiv 0$, which characterizes reparametrization invariant theories.
In that case the dynamics is described by $H_T = \sum v^{a_1} \Phi_{a_1}$,
where the sum extends only over the pimary first class constraints.
The Dirac algorithm ensures that these are in fact the only first class
constraints. Hence the gauge generator is described entirely in terms
of these constraints. As a result we obtain only
relations (\ref{multiplier}), and there are no restrictions 
on the infinitesimal gauge parameters. 

For some physically interesting models, the structure functions
$V^a_b$ and $C_{a,b}^c$ are actually constants, and the gauge parameters
are just functions of $t$. In the case where $C_{ab}^{c}=0$ and the
$V^a_b$'s are constant,    
equations (\ref{parameter}) have been solved
in ref. \cite{Shirzad2}. It leads to variations in the coordinates
which coincide with the general form given (\ref{deltaq}). 

An example where the $V^a_b$ depend on the coordinates is given by 
the pure Yang-Mills Lagrangian,
\be
{\cal L} = -\frac {1}{4} F^{a\mu\nu}F^a_{\mu\nu}\ .
\ee
This is a purely first class system with one primary constraint
\be
\pi^a_0(x)  \approx 0
\ee
and one secondary constraint
\be
[{\cal D}_i\pi_{i}]^a(x) = \partial_i\pi_i^a(x) + f_{abc}A^{b}_i(x)\pi_i^c(x)\approx 0\;.
\ee
The canonical Hamiltonian is given by
\be
H_c = \int d^3x \Big[\frac {1}{2} ({\vec\pi}^a)^2 + \frac {1}{4} (F^a_{ij})^2
+  A^a_0(D_i\pi_i)^a\Big]
\ee
>From this expression one readily finds that the non-vanishing structure 
functions analogous to those in (\ref{algebra1}) and (\ref{algebra})
 are given by
%%%%
\footnote{The index ``1" and ``2" refer to the primary and secondary constraint,
respectively.}
%%%%%
\be
(V^2_1(x,y))_{ab} = \delta (x-y)\delta_{ab}\ \ ; \ \ (V^2_2(x,y))_{ab} = 
gf_{acb}A^{0c}\delta(x-y)
\ee
\be
(C^2_{22}(x,y,z))_{abc} = -gf_{abc}\delta(x-y)\delta(y-z)
\ee
Hence in the case of the pure Yang-Mills theory  (\ref {parameter}) reduces to
\be
\partial_0\epsilon^a_2(x) = \epsilon^a_1(x) + gf_{acb}A^{0b}(x)\epsilon^c_2(x)
\ee
Taking $\epsilon^a_2(x) = \alpha^a(x)$ as the independent gauge parameters and 
solving the above equation for $\epsilon^a_1$ leads to the following 
structure of the gauge generator
\be
G(x) = ({\cal D}_0\alpha)^a(x)\pi^a_0(x) + ({\cal D}_i{\pi}_i)^a(x)\alpha^a(x)
\ee
>From this we immediately obtain for the infinitessimal gauge transformations 
of the potentials the familiar result, 
\be\label{deltagaugefield}
\delta A^a_\mu(x) = \int d^3y [A^a_\mu (x),G(y)] = ({\cal D}_\mu \alpha)^a(x)\;.
\ee
We can also compute the variation of the multiplier by using the first of
the Hamilton equations to obtain the relation $\dot A^a_0 = v^a$, and then 
using the commutativity analogous to (\ref{commutativity}). The result is, 
\be
\delta v^a = \partial_0 \delta A^a_0 = \partial_0 ({\cal D}_0\alpha)^a\;.
\ee
The same equation also follows directly from (\ref{multiplier})
with the identification $\epsilon^{b_1}\to \epsilon_1^a(x) = 
[{\cal D}_0\alpha(x)]^a$.

To complete our discussion we now reproduce the Lagrangian gauge identities
following from our Hamiltonian analysis. It is easy to see that the
variation (\ref{deltagaugefield}) can be cast into the form of (\ref{deltaq})
with $k$ taking the values $k=0,1$, where 
$\eta^{a_1}(t)$
are identified with with the gauge parameters $\alpha^a(x)$,
 and
\bear
\rho^{0a}_{(0)b}(x,y) &=& gf_{acb} A^{0c} \delta^3(x-y)\nonumber\\
\rho^{0a}_{(1)b}(x,y) &=& -\delta_{ab} \delta^3(x-y)\nonumber\\
\rho^{ia}_{(0)b}(x,y) &=& {\cal D}_{ab}^i \delta^3(x-y)\;.
\eear
Using these expressions in (\ref{gaugeidentities}), we arrive at
\be
\Lambda_a(x) \equiv ({\cal D}^\mu L_\mu)_a(x) = 0
\ee
where $L_\mu$ is the Euler derivative, which in the present case is given  by ${\cal D}^\sigma F_{\sigma\mu}$.
We have thus arrived at the standard gauge identity of the Yang-Mills
Lagrangian.

To conclude, we wish to emphasize once more the  main points of our
paper. We have derived for the most general case a master equation 
in the Hamiltonian formalism, which expresses the time independence of the
generator of gauge transformations,
and compactly encodes a pair of equations 
giving the restrictions on the
gauge parameters, as well as the variations of the Lagrange multipliers.
We have further explicitly demonstrated the consistency of this pair of
equations with Hamilton's equations of motion. The commutativity requirement 
(\ref{commutativity}) played a key role in 
the whole analysis. Observe that in a Lagrangian framework this
commutativity is always used when deriving the equations of motion, or
obtaining the gauge symmetries, while on the Hamiltonian level it
implies non-trivial restrictions on the gauge parameters.
The master equation was also shown to 
imply the invariance of
the total action, as well as the convariance of the Hamilton equations of motion.
We further discussed the connection with the purely Lagrangian approach.
In particular we established a correspondance between the gauge identities
and the master equation, which vanishes identically when expressed in terms of
the free parameters.

We have discussed  here systems involving only first class constraints.
The extension to mixed systems can be done in two distinct ways.
i) The conventional way would consist in replacing everywhere the
Poisson brackets by the corresponding Dirac brackets.
ii) An alternative procedure would consist in embedding the original theory
into a pure first class theory following the procedure of reference \cite{BFT}, and then to follow the steps given here. In particular,
purely second class systems could also be treated in this way.

\section*{Acknowledgement}

One of the authors (R.B.) would like to thank the Alexander von Humboldt Foundation for providing financial support making this collaboration
possible.

%%%%%%%%%%%%%%%%%%%%%%%%%%%%%%%%%%%%%%

\end{document}